\documentclass[reprint, aps, amsmath,amssymb]{revtex4-1} %% reprint,preprint
\usepackage{graphicx}
\usepackage{dcolumn}
\usepackage{bm}
\begin{document}

\title{Modified graphene with small BN domain: an effective method for band gap opening}

\author{Xiaofeng Fan}
\email{xffan@ntu.edu.sg} \affiliation{School of Physical and
Mathematical Sciences, Nanyang Technological University, 637616
Singapore} \affiliation{School of Materials Science and
Engineering, Jilin University, Changchun 130022, China}

\author{Jer-lai Kuo}
\affiliation{Institute of Atomic and Molecular Sciences, Academia
Sinica, Taipei 106, Taiwan
}%

\author{Zexiang Shen}
\affiliation{School of Physical and Mathematical Sciences, Nanyang
Technological University, 637616 Singapore
}%

\date{\today}

\begin{abstract}
BN domains are easy to form in the basal plane of graphene due to
phase separation. With first-principles DFT calculations, it is
demonstrated theoretically that the band gap of graphene can be
opened effectively around K (or K') points by doping the small BN
domains. It is also found that the random doping with B or N is
possible to open a small gap in the Dirac points, except for the
modulation of Fermi level. The surface charge which belongs to the
$\pi$ states near Dirac points is found to be redistributed
locally. The charge redistribution is contributed to the change of
localized potential due to the doping effects. This may be the
reason that the energy states near Dirac points is just disturbed
in the energy region [-0.4eV, 0.4eV] for the doping of BN domain.
Thus, we suggest that the band opening induced by the doped BN
domain is due to the breaking of localized symmetry of the
potential. Doping graphene with BN domains is an effective method
to open a band gap with quasi-linear dissipation near Dirac point.
\begin{description}
\item[PACS numbers]71.20.-b, 81.05.Uw, 71.10.-w
%\item[keywords]
\end{description}
\end{abstract}

\maketitle

\section{Introduction}
Graphene, consisted of a stable monolayer of carbon atoms with
honeycomb lattice, has attracted enormous
attention\cite{c1,c2,c3,c4} recently due to the fascinating
physical properties\cite{c5}, such as abnormal quantum Hall
effects\cite{c6} and massless Dirac fermions\cite{c7}, which is
attributed to the special linear behavior of electronic band near
Fermi level in $K$ (or $K^{'}$) points of Brillouin zone. However,
the linear gapless spectrum of graphene conflicts with the basic
requirement for its application in traditional electronic
devices\cite{c8, c9}, since in the most microelectronic devices,
such as field-effect transistors, the active material for the
transport layer needs the presence of energy gap by which the
conducting behavior of electrons or holes as charge carrier can be
controlled with the voltage. Therefore, with the expectation of
bandgap in graphene energy spectrum, the design of its band
structure has spurred an intense scientific interest. Many
different methods have been test to modulate the electronic
properties of graphene\cite{c10, c11,c12,c13, c14, c15,c16,c17}.

For engineering the band gap, these mechanisms tested can be
divided into three aspects. Firstly, with quantum limit effect of
size, the gap is possible to be opened by tweaking the width or
carving graphene into some geometry\cite{c12, c18}, such as
graphene nanoribbon. Secondly, it is to interact with different
substrates\cite{c19}, such as BN layer or [0001] SiC
substrate\cite{c11}, or to use electronic field with two-layer
graphene to break the symmetry of graphene for gap
opening\cite{c10, c14, c16}. Thirdly, modulating the electronic
properties is by functionalizing the graphene\cite{c15, c17, c20,
c21}. For examples, introducing Stone-Wales defects\cite{c13} in
the graphene layer, physically absorbing some chemical molecules
on graphene\cite{c20} and hydrogenation of graphene\cite{c17}.
Despite these recent progresses, an effective method for band gap
opening still remains to explore. It is noticed that the
difficulty of controlling the edge with scissor technique, the
complex of electrically gated bilayer graphene, the weak
disturbance to the electrons of graphene layer with physical
absorption and the destroy of planar network of graphene with
chemical absorbtion. It is also possible to be an effective method
that using traditional alloying mechanism to modulate the band gap
at $K$ ($K^{'}$) points of graphene. Unfortunately carbon stays
the second row of the table and it is difficult to find an
appropriate element to alloy the carbon with 2D structures.
Considering the compound BN is from the two elements near the
carbon, we will explore systematically the change of electronic
properties of graphene by alloying graphene with small
concentration of Boron nitride (BN).

BN is isoelectronic to carbon with similar structures, for
examples 3D structures cubic-BN and diamond, 2D layer structures
hexagonal-BN and graphite. It is known that BN sheet is with large
band gap. Thus, it is possible that the band gap will be opened
effectively by alloying graphene with BN. However, we have the
experience that BN is difficult to be solubilized with carbon for
any structures, such as, cubic phase, sheet structure and
fullerene-like structures. This is mostly contributed that the B-N
and C-C bonds with large bonding energy is stronger than the B-C
and N-C bonds and result in the phase separation with carbon
region and BN region. However, by properly controlling the growth
process, the BCN nanostructures with the rich variety of physical
properties can be synthesized\cite{c22,c22b}, such as the report
about single layer graphene with BN large domains by hybridizing
BN and graphene\cite{c22}. By the controlling of the thermodynamic
process or chemically introducing the hybridization of graphene
and BN molecule or $B_3N_3H_3$ molecule, the graphene with small
BN domains will be possible to be gained, since the lattice
constants of graphene is just a little difference from that of BN
sheet which makes the strain energy is small in the process of
alloying and the graphene fragment will be connected with small BN
domain in the thermodynamic process. In this paper, we will show
theoretically that the band gap of graphene can be opened
effectively around $K$ (or $K^{'}$) points with alloying the small
BN domains under the limit of low concentration, even under that
of the concentration for doping with BN domains.

With the density functional calculations, we explore firstly the
doping effect of B and N in graphene. With the unexpected case,
the doping effect on the 2D sheet graphene is different from that
on the 3D case of usual semiconductors. As expected, the B atom is
less an electron than C atom and the doping of B will result in
the down-shift of Fermi level for obtaining the hole as carrier.
The N atom is more an electron than C atoms and the doing of N
will induce the up-shift of Fermi level for gaining the electron
as carrier. Surprisingly, the B or N doping can open a gap around
Dirac point in the special case. Then we studied the question
carefully by analyzing the calculation. For the doping or alloying
of small BN domain, we detect the change of band structures and
the effect of band opening for the incorporation of BN molecule,
(BN)$_3$ and (BN)$_{12}$ for each concentration. With the analysis
of charge redistribution and potential redistribution due to the
introduction of small BN domain, it is found that the surface
charge states of graphene which belongs to the $\pi$ electrons is
sensitive to the redistribution. Thus, we suggest that the
redistribution of surface charge which breaks the local symmetry
of graphene is the main reason of band gap opening of graphene. In
additional, the coulomb dipole from BN molecule may be also
important to the breaking of symmetry of graphene structure and
thus to the band gap opening.

\section{Choice of Structures and Computational method}
\begin{figure}
\includegraphics [width=0.45\textwidth,clip] {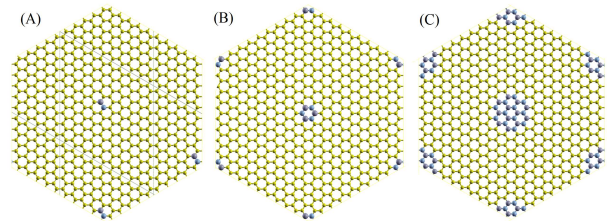}% Here is how to import EPS art%
\caption{\label{Fig-1}Schematic representation of graphene with
the doping of BN, (BN)$_3$ and (BN)$_{12}$ in $10\times10$ unit
cell to show that small BN domain is in graphene with low
concentration of BN. After the relaxation, the structures are not
obviously changed.}
\end{figure}

To model the 2D graphene with alloying or doping the B, N and
different BN domains, the supercell approach is used for the
structures. In order to avoid the spurious coupling effect along
$z$-axis due to the periodic images, the vacuum separation in the
model structures is set to 15 \AA. The graphene plane models are
constructed from hexagonal $2\times2, ~3\times3, ~4\times4,
~5\times5, ~6\times6, ~8\times8, ~10\times10$ and $12\times12$
unit cells, respectively. For alloying or doping the small BN
domains, we consider the BN molecule, (BN)$_{3}$ and (BN)$_{12}$
to merge into the graphene basal plane, as shown in
Fig.\ref{Fig-1}.

In the present work, the accurate frozen-core full-potential
projector augmented-wave (PAW) method, as implemented in the
Vienna $ab$ initio simulation package (VASP)\cite{c23, c24} is
used. With density functional theory, the electronic density with
the exchange and correlation effects is computed under the local
density approximation (LDA). The k-space integral and plane-wave
basis, as detailed below, are tested to ensure the total energy is
converged at the 1 meV/atom level. The kinetic energy cutoff of
550 eV for the plane wave expansion is found to be sufficient and
the Monkhorst-Pack method is used to choose k points in moment
space. By the optimization process, the lattice constants of
graphene and BN sheets obtained are 2.44\AA~ and 2.48\AA,
respectively. The values are near the experimental value obtained
at low temperature (2.46\AA~for graphene and 2.50\AA~for h-BN)
with small underestimation. It is noticed that the electronic
bands near surface is not very sensitive to the small change of
lattice constant for the graphene and single layer BN sheet.
Therefore, for the optimization of the systems with doping or
alloying with low concentration, the lattice parameters are fixed
with the consideration of Vegard's law and the internal
coordinates of atoms are relaxed fully. After relaxation of
structures, the strain on the systems is very small and the
relaxed structures are found to follow Vegard's rules well.

\section{Results and Discussion}
\subsection{Doping pristine graphene by B and N}
Due to the weaker bond energy of B-B and N-N than C-N and C-B, the
boron or nitrogen is relatively easy to alloy with $sp_2$-
hybridized carbon materials, such as graphite, CNT, graphene oxide
and graphene. With the relaxation and optimization of the
structure, it is found that the lattice constant just has a little
change and the local strain is also small due to the introduction
of B or N in the lattice of graphene as a doping effect. This
means that the 2D planar structure is not destroyed due to the
small amount of B, N and BN which is not like the effect of
chemical absorption, such as hydrogen absorbed on graphene.

\begin{figure}
\includegraphics [width=0.32\textwidth,clip] {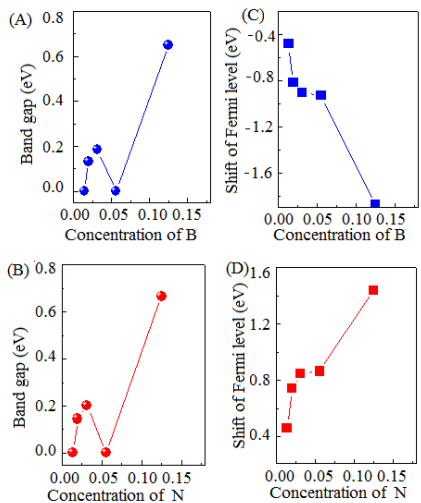}% Here is how to import EPS art%
\caption{\label{Fig-2} The band gap (A) and shift of Fermi level
(C) as a function of concentration of boron in graphene and the
band gap (B) and shift of Fermi level (D) as a function of
concentration of nitrogen in graphene. Notice that the structure
of each concentration calculated is with the unit cell $6\times6$,
$5\times5$, $4\times4$, $3\times3$ and $2\times2$, respectively.}
\end{figure}

For the usual semiconductors, the point defect as an acceptor or
donor is introduced into the lattice of host will induce an
accepter energy level near the top of valance band or donor energy
level near the bottom of conduction band in the band gap of the
host's energy spectrum and thus the formation of the $p$-type or
$n$-type material. However, Duo to the zero band gap of graphene,
it is possible that boron or nitrogen in the lattice matrix of
graphene will result in the special defect energy level in valance
bands or conduction bands which are contributed to the $p$
orbitals. As shown in Fig.\ref{Fig-2} C and D, the doping of B
results in the down-shift of Fermi level and the doping of N
results in the up-shift of Fermi level. With the thermal anneal of
graphene ribbon in ammonia, the experiment have confirmed that the
introduction of nitrogen can result in $n$-doped graphene
material\cite{c25}.

Near Dirac points, the linear dispersion can be expressed as
$E_{\pm}\approx\pm\upsilon_F|q|$, where $q$ is the momentum
related to the Dirac points and $\upsilon_F$ is Fermi
velocity\cite{c2}. With the linear relation, the density of states
per unit cell can be given by $\rho=2 A |E|\pi \upsilon_F^2$,
where $A = \sqrt{3}a^2/2$ is the unit cell area with the lattice
constant $a$=2.42\AA. Thus, with the doping of small amount of
dopant, the shift of Fermi level is related with the concentration
of dopant by the relation $\Delta E_F\propto \sqrt{n}$ (1), where
$n$ is the concentration of boron or nitrogen. In the low
concentration, the doping effect follows the tendency of formula
(1), as shown in Fig. \ref{Fig-2} C and D.

\begin{figure}
\includegraphics [width=0.4\textwidth,clip] {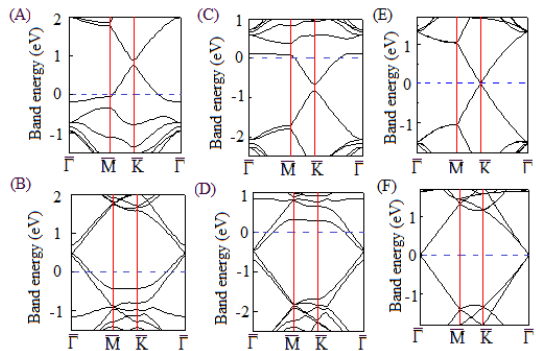}% Here is how to import EPS art%
\caption{\label{Fig-3}The band structures of single B-doped
graphene with $5\times5$ unit cell (A) and $6\times6$ unit
cell(B), and of single N-doped graphene with $5\times5$ unit cell
(C) and $6\times6$ unit cell(D). The band structures of graphene
with $5\times5$ unit cell (E) and $6\times6$ unit cell (F) given
as a reference.}
\end{figure}

Though the boron or nitrogen in the lattice matrix of graphene
does not induce any defect energy states near Dirac points, the
band states near the Dirac points have been disturbed to some
extent. As shown in Fig. \ref{Fig-2} A and B, in some
concentration, an energy gap is induced around the Dirac point. It
may be considered that the symmetry of the two sublattices is
broken by the doped boron or nitrogen is the important reason.
However, this can not explain that the band gap is not opened in
the B or N-doped graphene with $3n\times3n$ ($n=1, 2, 3,
\cdot\cdot\cdot$) unit cell. We need to check the change of band
structure of graphene due to the introduction of B or N. In Fig.
\ref{Fig-3}, we compare the band structure of B- or N-doped
graphenes to the prime graphenes with $5\times5$ and $6\times6$
unit cells. It is assured that there are not the localized defect
states in the energy spectrum of doped graphene. This may be
attributed to the strong interaction of B (or N) and C and the
small difference of electronegativity, with the
$sp_2$-hybridization. For B-doped graphene, a new hybridized band
around -1.0 $\sim$ -1.5 eV under Dirac point is formed. For
N-doped graphene, there is a new hybridized band at about 1.0 eV
above the Dirac point. The new band affects strongly the band
structure near energy ranges and thus the charge will be possible
to be redistributed (see the subsection below for details).
Therefore, the bands near the Dirac points will be disturbed. For
the lattice with $3n\times3n$ unit cell, the Dirac points are
folded to the $\Gamma$ point due to the reduction of first
Brillouin zone. It is well known that the two carbon in the unit
cell which is not equivalent from symmetry result in the symmetry
non-equivalence of $K$ and $K'$ points in the momentum space. The
states around Dirac points from the $K$ and $K'$ points will
interact at the $\Gamma$ point of reduced Brillouin zone for B or
N-doped graphene with $3n\times3n$ unit cell. Therefore, the
states near Dirac point from different bands are hybridized and
thus the band gap is closed.

\subsection{Doping pristine graphene with small BN domains}
\begin{figure}
\includegraphics [width=0.4\textwidth,clip] {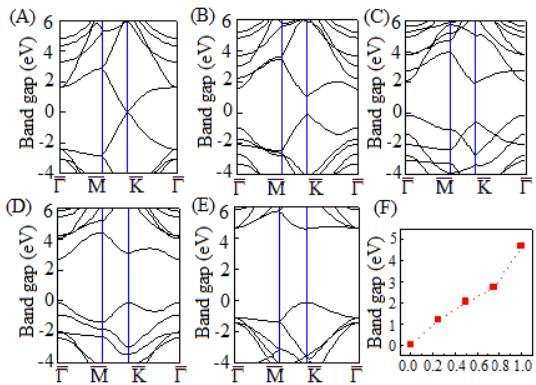}% Here is how to import EPS art%
\caption{\label{Fig-4}The band structures of graphene (A),
$(BN)_1C_6$ (B), $(BN)_2C_4$ (C), $(BN)_3C_2$ (D) and $BN$ (E)
calculated in $2\times2$ unit cell, and the band gap as a function
of the concentration BN (F).}
\end{figure}

We firstly analyze the band structure of alloy
$(BN)_{1-x}(C_2)_x$. It is noticed that the B and N domain is
impossible to be formed in the alloy due to the weaker binding
energy of B-B and N-N bonds in the planar structure. Thus, we
simply construct the alloy structure in $2\times2$ unit cell
($(BN)_1C_6$ and $(BN)_3C_2$ with symmetry group $AMM_2$,
$(BN)_2C_4$ with symmetry group $PMM_2$) and the band structures
are calculated, as shown in Fig.\ref{Fig-4}. The Fermi level is
found to be not shifted and the two bands near the Dirac point is
the major characteristic for the alloys. For the concentration of
BN is 50$\%$ and 75$\%$, the band gap is indirect. With the
concentration 50$\%$, the energy difference of $T\rightarrow K$,
$T \rightarrow T$ and $K \rightarrow K$ are 2.058 eV, 2.24 eV and
2.48 eV, respectively. With the concentration 75$\%$, the
transition energies of $K \rightarrow T$, $T \rightarrow T$, $K
\rightarrow K$ are 2.73 eV, 2.75 eV and 3.235 eV, respectively.
The relation between band gap and concentration is almost linear
with a small concavity around the concentration 75$\%$, as shown
in Fig. 4F. Therefore, with alloying BN, the band structure can be
modified effectively with the opened band gap. However, It is
obvious that the ordered alloy $(BN)_{1-x}(C_2)_x$ with 2D planar
structure is difficult to form, since BN just have small
solubility in carbon materials.

\begin{figure}
\includegraphics [width=0.4\textwidth,clip] {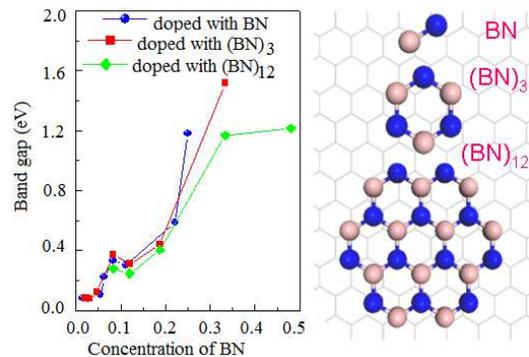}% Here is how to import EPS art%
\caption{\label{Fig-5}The band gap as a function of concentration
BN for the low-concentration doping of graphene with the models in
which the cluster region is BN molecule, (BN)$_3$ and (BN)$_{12}$,
respectively.}
\end{figure}

With the thermodynamic process, it is possible that small amount
of BN can be doped into the planar lattice of graphene. Due to the
separation of phase, BN will precipitate and form the domain in
the graphene. With the thermal catalytic CVD method, Ci $et~al.$
have gained the hybrid h-BNC material with localized BN domains
and graphene domains\cite{c22}. By the development of material
growth techniques, it is possible to obtain the graphene with
small localized domain of BN. Now we considered what the shape of
localized domains is. Since the binding energies of C-C and B-N
bonds are larger than that of B-C and N-C bonds, the B-N and C-C
bonds will tend to segregate and thus that the relative stable
states will be with smaller amount of B-C and N-C bonds. With
consideration of bond rule, the sum of the number of C-C, B-N, B-C
and N-C is a constant. For the localized regions with different
shapes and same areas, the domain with circle will have the
smallest length of the boundary and thus the smallest number of
B-C and N-C bonds. Therefore, the small domain of BN, (BN)$_{3}$
and (BN)$_{12}$ will be possible to exist in the graphene with the
shapes which is shown in Fig. \ref{Fig-5}. In additional, it is
possible that the doped graphene can be obtained by the
combination of small pieces of graphene (such as $C_6H_6$) and BN
molecules (or $(BN)_3H_3$, $(BN)_{12}H_6$) with chemical methods.

\begin{figure}
\includegraphics [width=0.4\textwidth,clip] {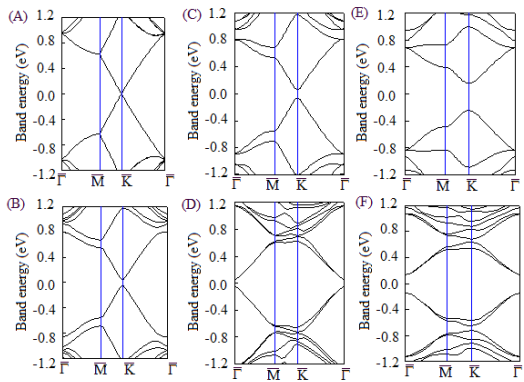}% Here is how to import EPS art%
\caption{\label{Fig-6}The band structures of graphene (A) with
$8\times8$ unit cell and BN-doped graphene (B) for the model of
graphene with BN molecule in $8\times8$ unit cell, (BN)$_3$-doped
graphene for the model of graphene with (BN)$_3$ cluster in
$8\times8$ unite cell (C) and $12\times12$ unite cell (D),
(BN)$_{12}$-doped graphene for the model of graphene with
(BN)$_{12}$ cluster in $8\times8$ unite cell (E) and $12\times12$
unite cell (F), respectively.}
\end{figure}

Here we demonstrate the electronic properties of doped graphene
with small domain of BN, by calculating the doping of BN,
(BN)$_{3}$ and (BN)$_{12}$ with different unit cell for the
consideration of domain size and the concentration of dopants, as
shown in Fig. \ref{Fig-5} and Fig. \ref{Fig-6}. With the BN
doping, the band gap is opened in the Dirac points and Fermi level
is still around the Dirac points and not shifted. Due to the
isoelectronic property of BN and C$_2$, the codoping of B and N is
similar with the alloying effect of BN as shown in Fig.
\ref{Fig-4} and the localized defect stats is not found in the
band structure, as shown Fig.6. However, the change of band gap
with the increase of concentration due to the doping of BN domains
is not linear and there are two dips around the concentrations
0.04 and 0.12 in the curves of band gap vs. concentration. In the
low concentration region, the charge of band gap due to the
difference of domain size is small. This means the band gap in the
Dirac points can be modulated effectively with the formation of
small BN domain in the lattice of graphene with small
concentration.

In Fig. \ref{Fig-6} A and B, the band structure of graphene and
that doped by BN molecule with $8\times8$ unit cell is shown. It
can be found that the bands of both structures match perfectly in
the energy region [-0.4, 0.4] except the small region around the
Dirac points. In the higher or lower energy region, the
introduction of BN in lattice of graphene has broken the symmetry
of band states and given an obvious perturbation to the eigenvalue
of each state to make the small up- or down-shift for the energy.
It is obvious that the conical dispersion is retained and the band
departs from the linear dispersion asymptotically just around the
Dirac points and thus opens a gap in the small region. With the
increase of the concentration, the linear dispersion is modified
in a relatively larger region of energy with a lager band gap. The
phenomenon that bands around Dirac points are modulated is not
changed obviously following the change of size of the small BN
domains, as shown in Fig. \ref{Fig-6} C, D, E and F. It is
interesting that in the case of B or N doping with $3n\times3n$
unit cell, the eigenstates around the Dirac points is obviously
perturbed and the degenerated energies of $\pi$ and $\pi^*$ states
are destroyed with zero band gap due to the symmetry break of the
two sublattices. In the case of BN codoping, the degenerated
energies of $\pi$ or $\pi^*$ states are not broken with an opened
band gap. This may be due to the isoelectronic property of BN for
C$_2$ and the redistribution of charge in the region near the
Dirac points.

\subsection{Distribution of charge density and local potential of doped graphene}
\begin{figure}
\includegraphics [width=0.4\textwidth,clip] {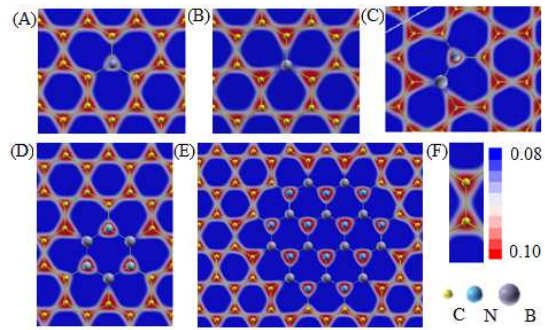}% Here is how to import EPS art%
\caption{\label{Fig-7}The distribution of charge density in $xy$
plane for the region at about 1.2\AA~ above/under the graphene
plane along $z$ direction for N-doped graphene (A), B-doped
graphene (B), graphene with BN molecule (C), graphene with
(BN)$_{3}$ cluster (D), graphene with (BN)$_{12}$ cluster (E) and
pure graphene (F), respectively.}
\end{figure}

It is well known, with $sp_2$ hybridization, the stable 2D
hexagonal sheet is formed with the $\sigma$ bonds. Obviously, the
$\sigma$ bond is strong can be contributed that the $sp_2$
hybridized orbitals of the nearest-neighbour carbon atoms is
superposed largely and thus makes the charges are mostly
distributed in the region between two carbons. Duo to Pauli
exclusion principle, the electron of the remnant $p_z$ orbitals
will be distributed on the surface of network of $\sigma$ bonds.
This results that the electronic states of $\pi$ orbitals are
sensitive to the local perturbations and the electrical properties
of graphene is easy to be changed after the physical or chemical
adsorption of gaps molecules due to the charge exchange and charge
transfer. It is interesting that the electron charges of $\pi$
orbitals near the Dirac points are mostly distributed on the
carbon atoms due to the electrons of $\sigma$ mostly at the region
between two carbons, as shown in Fig. \ref{Fig-7} F. This is also
the reason that the electronic states are put on the carbon atom
in the theoretical model which is used to study the low-energy
electronic properties near the Dirac points.

With the introduction of B, N or BN into the lattice of graphene,
the charge distribution of $\sigma$ bonds is not obviously
changed. At the same time, the distribution of $\pi$ electrons is
disturbed strongly. As Fig. \ref{Fig-7} A shown, the charge of
nitrogen near the Dirac point is localized and isolated and the
charge of three neighbor carbons is reduced. It is obviously that
some of charges of N atom is transferred into the electronic
states of $\pi^*$ orbitals of carbon atoms and thus make the Fermi
level up-shift. For B-doping, as shown in Fig. \ref{Fig-7} B, the
charge from other carbon has a trend to congregate to B atom
because the B atom is less one electron than C atom. Thus, this
makes the charge of three neighbor carbon atoms increased and the
localized congregation of charge around boron atom makes the Fermi
level down-shift. The projected charge distribution of doped
graphene with BN molecule, (BN)$_3$ and (BN)$_{12}$ domains are
demonstrated in Fig. \ref{Fig-7} C, D and E, respectively. It is
clear that the effect of B and N in the BN codoping to the
neighbor carbon atoms is very similar to that of the isolated B-
or N-doping. The effect of domain of BN to the carbon atoms of the
sheet plane is localized. Due to isoelectronic property of BN and
C$_2$, the Fermi level is not shifted. Thus, the open of band gap
may be attributed to the localized symmetry break of sublattices.

\begin{figure}
\includegraphics [width=0.4\textwidth,clip] {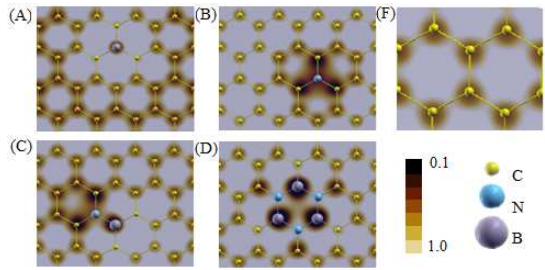}% Here is how to import EPS art%
\caption{\label{Fig-8}The local potential in $xy$ plane in $xy$
plane for the region at about 1.2\AA~ above/under the graphene
plane along $z$ direction for B-doped graphene (A), N-doped
graphene (N), graphene with BN molecule (C), graphene with
(BN)$_3$ cluster (D), pure graphene (F), respectively.}
\end{figure}

As shown in Fig. \ref{Fig-8} F, the local potential in the plane
$xy$ paralleled to the sheet plane with a distance about 1.2\AA~
reveals the reason that the charge near Dirac points localized on
carbon atoms. With the B-doping shown in Fig. \ref{Fig-8}A, the B
atom has a localized low potential and thus attracts the other
electron from carbon atoms. The attraction is not enough that
makes the redundant charge localized on B atom. However, this
results in the redundant charge localized on three neighbor
carbons. For the N-doing, as show in Fig. \ref{Fig-8} B, the
potential of three carbons near boron is lower. Thus, the charge
of N atom has a trend to diffuse to the near carbon atoms. As
shown Fig. \ref{Fig-8} C and D, the introduction of BN domain in
carbon lattice breaks the symmetry of potential which decides the
distribution of charge near the Dirac points.

\section{Conclusion}
In summary, the effects of B-, N- and BN-doping or alloying to the
electronic property has been studied by first-principles DFT
calculations. It is found that the introduction of small BN domain
in the lattice of graphene can be an effective method to open the
band gap at the $K$ (or $K^{'}$) points. The modulation of gap is
not sensitive to the size of BN domain and dependent on the
concentration of BN. For B or N doping, it is found that the shift
of Fermi level depends on the concentration of B or N. The doping
of B or N can also open a gap in the Dirac points, while it is
found that the single atom doping is not an effective method for
engineering band structure of graphene. Furthermore, the band gap
can not be opened when the graphene with $3n\times3n$ unit cell is
doped by B or N. This can be contributed that the perturbed
potential result in the coupling of $\pi$ and $\pi^*$ states near
Dirac points from $K$ and $K^{'}$ both which is folded to the
gamma point. However, we can predict that the band gap of the
actual sample with B or N doping should be opened due to the
disorder effect of doping.

By analyzing the projected charge distribution, the surface charge
which belongs to the $\pi$ states near the Dirac points is
obviously disturbed and redistributed after the doping of B, N or
BN domains, through the electronic states belong to the $\sigma$
bands are not visually changed. The charge redistribution due to
the doping is localized may be the reason that the energy states
near Dirac points is just disturbed in the energy region [-0.4,
0.4] for the doping of BN domain. With the analysis of projected
potential, the charge redistribution is considered to be a result
of the small change of localized potential due to the doped
defects. Thus, the band opening due to the doping of BN domain can
be attributed to the breaking of localized symmetry. From the
charge redistribution, we also suggest that the scattering from
the doped defects may have not evident effect to the mobility of
carriers in the graphene. With the doping of BN domain, we can
obtain an effective band gap for the application of graphene on
next-generation microelectronic devices.

%\clearpage

%\newpage

%\bibliography{basename of .bib file}

\begin{thebibliography}
\bibitem{}
\bibitem{c1}
K. S. Novoselov, A. K. Geim, S. V. Morozov, D. Jiang, Y. Zhang, S.
V. Dubonos, I. V. Grigorieva, and A. A. Firsov, Science {\bf306},
666 (2004).
\bibitem{c2}
A. H. Castro Neto, F. Guinea, N. M. R. Peres, K. S. Novoselov, and
A. K. Geim, Reviews of Modern Physics {\bf81}, 109 (2009).
\bibitem{c3}
A. K. Geim and K. S. Novoselov, Nature Materials {\bf6}, 183
(2007).
\bibitem{c4}
M. J. Allen, V. C. Tung, and R. B. Kaner, Chemical Reviews
{\bf110}, 132 (2009).
\bibitem{c5}
S. V. Morozov, Phys. Rev. Lett. {\bf97}, 016801 (2006).
\bibitem{c6}
Y. Zhang, J. W. Tan, H. L. Stormer, and P. Kim, Nature {\bf438},
201 (2005).
\bibitem{c7}
K. S. Novoselov, A. K. Geim, S. V. Morozov, D. Jiang, M. I.
Katsnelson, I. V. Grigorieva, S. V. Dubonos, and A. A. Firsov,
Nature {\bf438}, 197 (2005).
\bibitem{c8}
K. Novoselov, Nat Mater {\bf6}, 720 (2007).
\bibitem{c9}
P. Avouris, Z. H. Chen, and V. Perebeinos, Nature Nanotechnology
{\bf2}, 605 (2007).
\bibitem{c10}
T. Ohta, A. Bostwick, T. Seyller, K. Horn, and E. Rotenberg,
Science {\bf313}, 951 (2006).
\bibitem{c11}
S. Y. Zhou, G. H. Gweon, A. V. Fedorov, P. N. First, W. A. de
Heer, D. H. Lee, F. Guinea, A. H. Castro Neto, and A. Lanzara, Nat
Mater {\bf6}, 770 (2007).
\bibitem{c12}
M. Y. Han, Ouml, B. zyilmaz, Y. Zhang, and P. Kim, Physical Review
Letters {\bf98}, 206805 (2007).
\bibitem{c13}
X. Peng and R. Ahuja, Nano Letters {\bf8}, 4464 (2008).
\bibitem{c14}
J. B. Oostinga, H. B. Heersche, X. Liu, A. F. Morpurgo, and L. M.
K. Vandersypen, Nat Mater {\bf7}, 151 (2008).
\bibitem{c15}
D. W. Boukhvalov and M. I. Katsnelson, Physical Review B {\bf78},
085413 (2008).
\bibitem{c16}
Y. Zhang, T.-T. Tang, C. Girit, Z. Hao, M. C. Martin, A. Zettl, M.
F. Crommie, Y. R. Shen, and F. Wang, Nature {\bf459}, 820 (2009).
\bibitem{c17}
D. C. Elias, et al., Science {\bf323}, 610 (2009).
\bibitem{c18}
A. K. Singh, E. S. Penev, and B. I. Yakobson, ACS Nano {\bf4},
3510 (2010).
\bibitem{c19}
G. Giovannetti, P. A. Khomyakov, G. Brocks, P. J. Kelly, and J.
van den Brink, Physical Review B {\bf76}, 073103 (2007).
\bibitem{c20}
X.F. Fan, L. Liu, J.-L. Kuo, and Z.X. Shen, The Journal of
Physical Chemistry C {\bf114}, 14939 (2010).
\bibitem{c21}
T. O. Wehling, K. S. Novoselov, S. V. Morozov, E. E. Vdovin, M. I.
Katsnelson, A. K. Geim, and A. I. Lichtenstein, Nano Letters
{\bf8}, 173 (2007).
\bibitem{c22}
L. Ci, et al., Nat Mater {\bf9}, 430 (2010).
\bibitem{c22b}
A. Rubio, {\bf9}, 379¨C380 (2010)
\bibitem{c23}
G. Kresse and J. Furthm¨¹ller, Computat. Mater. Sci. {\bf6}, 15
(1996).
\bibitem{c24}
G. Kresse and J. Furthm¨¹ller, Phys. Rev. B {\bf54}, 11169 (1996).
\bibitem{c25}
X. Wang, X. Li, L. Zhang, Y. Yoon, P. K. Weber, H. Wang, J. Guo,
and H. Dai, Science {\bf324}, 768 (2009).
\end{thebibliography}

%\clearpage

%\newpage

\end{document}